\documentclass[letterpaper, 10 pt, conference]{ieeeconf}  

\IEEEoverridecommandlockouts                              

\usepackage{graphics} 
\usepackage{times} 
\usepackage{amsmath} 
\usepackage{amssymb}  
\usepackage{bbold}
\usepackage{arydshln}
\usepackage{multirow}
\usepackage{booktabs} 

\usepackage{adjustbox}
\usepackage{svg}
\usepackage{floatrow}
\usepackage{amsmath}
\usepackage{balance}

\usepackage{blindtext}

\usepackage{xcolor}
\usepackage{soul}

\usepackage[nobiblatex]{xurl}

\usepackage{physics}

\usepackage{pifont}

\usepackage{adjustbox}

\usepackage{hyperref}

\usepackage{graphicx}

\let\labelindent\IEEElabelindent
\usepackage{enumitem}

\definecolor{fei}{rgb}{0.0, 0.0, 0.0}

\title{\LARGE \bf
HemoSet: The First Blood Segmentation Dataset for Automation of Hemostasis Management
}

\author{Albert J. Miao$^1$, Shan Lin$^1$, Jingpei Lu$^1$, Florian Richter$^1$, \\Benjamin Ostrander$^2$, Emily K. Funk$^2$, Ryan K. Orosco$^3$, Michael C. Yip$^1$, \textit{Senior Member, IEEE} 
\thanks{{}$^1$A.J. Miao, S. Lin, J. Lu, F. Richter, and M.C. Yip are with the Department of Electrical and Computer Engineering, University of California San Diego, La Jolla, CA 92093, USA.
        (e-mail: \{amiao, shl102, jil360, frichter, yip\}@ucsd.edu)}
\thanks{{}$^2$B. Ostrander and E. Funk are with the Department of Otolaryngology, University of California San Diego, La Jolla, CA 92093, USA.}
\thanks{{}$^3$R. Orosco is with the Department of Surgery, University of New Mexico, Albuquerque, NM 87102, USA.}}

\begin{document}
\bstctlcite{IEEEexample:BSTcontrol}

\maketitle
\thispagestyle{empty}
\pagestyle{empty}

\begin{abstract}
Hemorrhaging occurs in surgeries of all types, forcing surgeons to quickly adapt to the visual interference that results from blood rapidly filling the surgical field.
Introducing automation into the crucial surgical task of hemostasis management would offload mental and physical tasks from the surgeon and surgical assistants while simultaneously increasing the efficiency and safety of the operation.
The first step in automation of hemostasis management is detection of blood in the surgical field.
To propel the development of blood detection algorithms in surgeries, we present HemoSet, the first blood segmentation dataset based on bleeding during a live animal robotic surgery.
Our dataset features vessel hemorrhage scenarios where turbulent flow leads to abnormal pooling geometries in surgical fields.
These pools are formed in conditions endemic to surgical procedures --- uneven heterogeneous tissue, under glossy lighting conditions and rapid tool movement.
We benchmark several state-of-the-art segmentation models and provide insight into the difficulties specific to blood detection.
We intend for HemoSet to spur development of autonomous blood suction tools by providing a platform for training and refining blood segmentation models, addressing the precision needed for such robotics.
\footnote{Our dataset and code are available at \url{https://arclab-hemoset.github.io}}

\end{abstract}


\section{Introduction} \label{sec:intro}


Automation for surgery has progressed across multiple disciplines to tackle the wide variety of surgical tasks including Augmented Reality (AR) for intuitive visual cues \cite{vavra2017recent}, robotic systems that surgeons operate with \cite{yip2019robot}, and video analysis for post-operation review \cite{kawka2022intraoperative}.
Even with the significant progress found in automation for surgery, the crucial surgical task of hemostasis management, the process of managing and controlling bleeding during surgery such that there are no occlusions from blood, has not received much attention.
Hemorrhaging occurs in surgeries of all types, forcing surgeons to quickly adapt to the visual interference that results from blood rapidly filling the surgical field and establishing hemostasis as quickly as possible to minimize bleeding and allow the critical tasks of the operation to proceed.

\begin{figure}[t!]
\includegraphics[width=\textwidth]{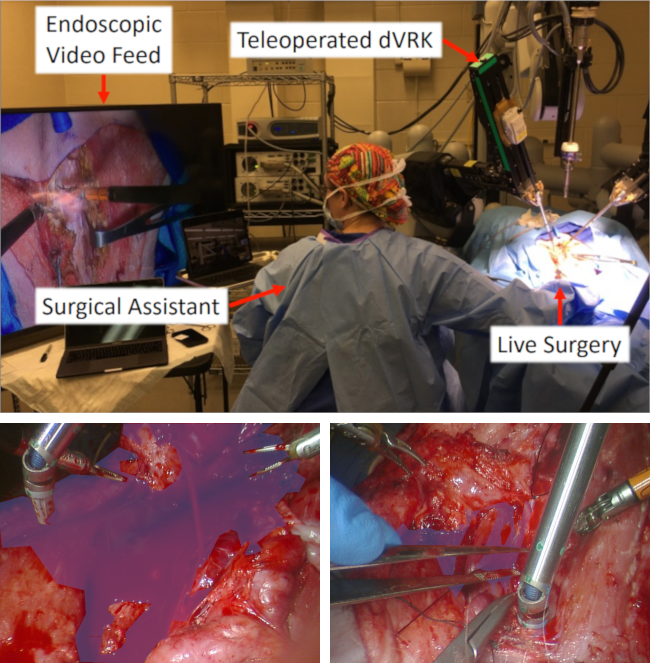}
\caption{HemoSet is collected from a thyroidectomy performed on a porcine model.
The dataset is collected by identifying blood vessels to induce hemorrhaging during the procedure.
A layout of the procedure (top) and example labeled images from the dataset (bottom).
} \label{fig1}
\end{figure}

Significant instrumentation and tools have been designed for surgeons to control bleeding (e.g. cautery, staplers, sutures), since the surgical task of maintaining hemostasis is universal to all surgeries.
However, automation efforts in hemostasis management have been limited to only a few works, all of which directly avoid the challenge of segmenting blood from surgical cameras via image classification (blood or no blood), namely event-based detection of active bleeding and automation of the suction tool to remove blood \cite{marullo2023multi}, hand-crafted optical flow filtering \cite{richter2021autonomous}, color-based detection in a lab bench-top setting \cite{huang2021model}, and over-fitting to the specific lab bench-top setting \cite{noguera2021controlling}.

\begin{figure}[t!]
    \centering
    \makebox[1\textwidth][c]{\includegraphics[width=1.05\textwidth]{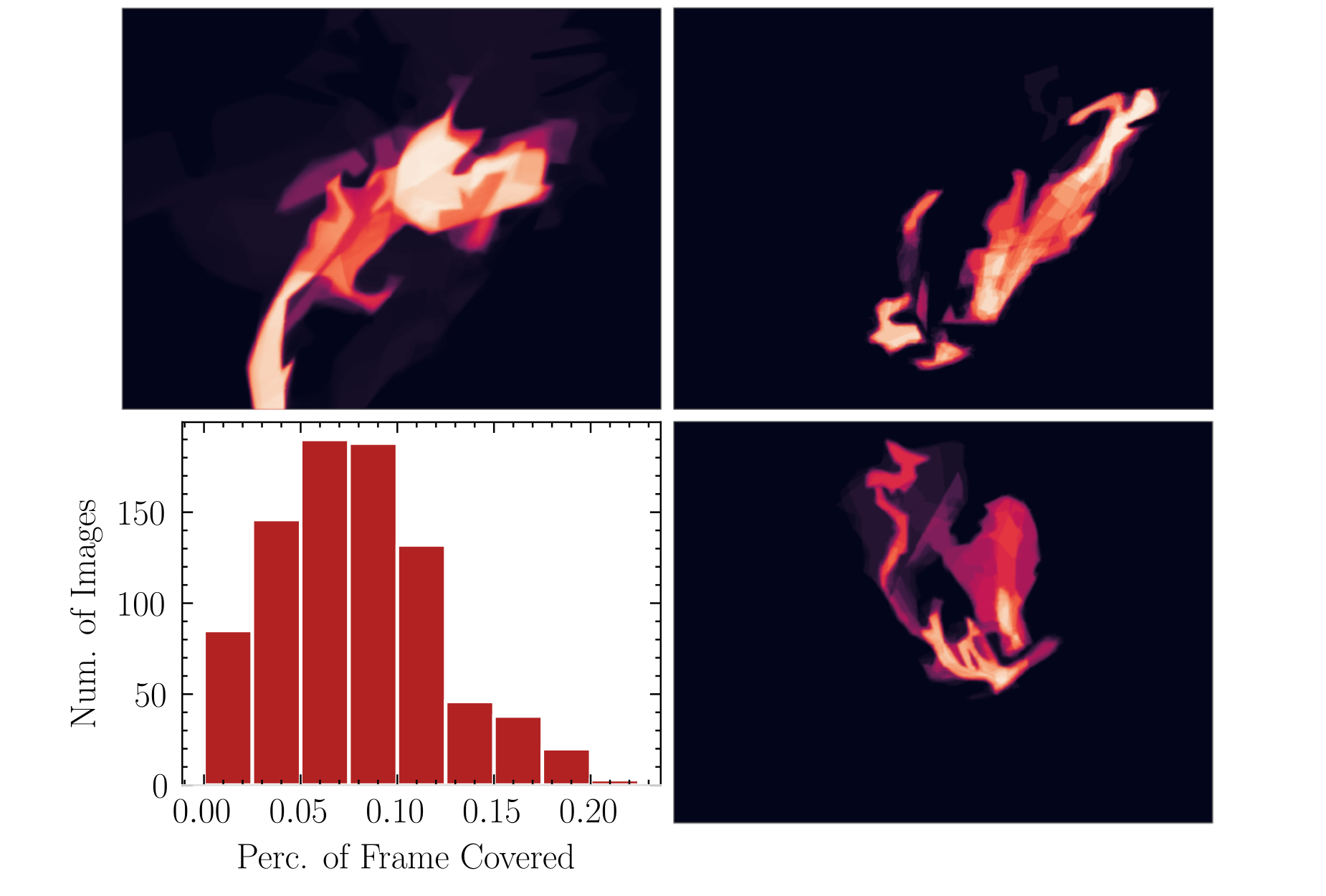}}
    \caption{The percentage of blood coverage over labeled frames alongside heatmaps of blood across trials 6, 10, and 11 (without augmentation). Our dataset is distinct from previous works in the high variation of blood coverage. The controlled environment of previous datasets leads to poor generalization when moving towards real-world applicability. Our dataset has the variety in blood geometry necessary to train a practical blood-detection algorithm.}
    \label{fig:datastat}
\end{figure}

\begin{table}[t]
    \caption{Overall HemoSet Statistics}\label{tab:datastat}
    \begin{tabular}{l@{\hskip3cm}l}
        \hline
        \textbf{HemoSet} & \\
        \hline
        Videos & 11 \\ 
        Frames & 102,616 \\
        Labeled Frames & 857 \\
        FPS & 30 \\
        Resolution & 640$\times$480 \\
        Avg. Image Coverage & 8.06\% \\
        Avg. Duration & 2m37s \\
        Avg. STAPLE Prec. & 93.3\% \\
        Avg. STAPLE Spec. & 99.6\% \\
        \hline
    \end{tabular}
\end{table}

These works stress the importance of hemostasis management, but general segmentation algorithms, such as Meta's Segment Anything Model (SAM) \cite{kirillov2023segment} are unable to adapt to either the visual context of a surgical scene or the geometry of pooled blood, shown in Figure \ref{fig:sam_comparison}. However, training a dedicated blood segmentation algorithm is severely limited by the absence of a reliable dataset, which is essential for bridging the gap between the problem and the real solution.
Given a sufficient dataset, other automation methods can be considered such as AR guidance for managing hemostasis, post-operative video analysis on hemorrhages, and even estimation of blood loss from the patient.




In this work, we propose the HemoSet, the first to feature segmented blood in a surgical setting, addressing the need for labeled data that is representative of the domain. We highlight the key challenges of this dataset as follows: \textbf{1) erratic pooling geometry:} the turbulent nature of the environment causes unique contours with each incision, \textbf{2) distinction of pools and stains:} they are categorically distinct relative to hemostasis despite their visual similarity, and \textbf{3) varying difficulty:} the variety of incisions and pool contours ensure novel difficulty levels across different segments of the video.
We believe that our dataset will be a valuable resource for researchers and developers working on automatic hemostasis algorithms and systems.


We further benchmark the performance of existing segmentation algorithms on our dataset and provide recommendations for future research in this area. Namely, 1. a blood segmentation model should prioritize the preservation of initial low-level features in order to accurately label edges and avoid annotating stains, and 2. the varying difficulty must be kept in mind when evaluating model comparisons. We hope that our dataset will stimulate further research and development in automated hemostasis, leading to more effective and efficient surgical procedures.

\section{Related Work}


The progress of automated segmentation in endoscopy is observed through the quality and variety of labeled datasets and the performance of existing models on them. Yet, research on blood detection and segmentation remains under-explored primarily due to limited access to annotated datasets for blood segmentation, and its adaptive nature (e.g. varying blood pool shapes in different surgical scenarios), posing unique challenges for existing algorithms. 

\textbf{Liquid Segmentation} is a niche field among computer vision tasks \cite{schenck2018perceiving,yamaguchi2016stereo}, given the unusual and varying shapes fluids can form \cite{richter2022image}. We see such datasets in medicine with microvascular detection \cite{gur2019unsupervised,haft2019deep} or Wireless Capsule Endoscopy (WCE) \cite{karargyris2011detection}, but these papers focus on clinical diagnosis rather than intervention, making them entirely distinct from hemorrhage segmentation. Hemorrhage segmentation datasets must encompass the challenges endemic to live surgery in order to best represent the domain, especially when considering automation tasks \cite{richter2021autonomous}. We acknowledge current examples of blood detection in surgical scenes \cite{marullo2023multi,noguera2021controlling,tang2022bleeding}, but they circumvent the challenges associated with hemorrhage segmentation by using data unrepresentative of a live surgery. 




\textbf{Segmentation Algorithms} are similarly varied, and modern techniques rely on convolutional neural networks (CNNs), attention-based mechanisms, transformers, and combinations thereof to extract patterns of features in the image. General segmentation models such as Meta's Segment Anything model have shown strong performance on in-the-wild data but lack the fidelity to adapt to the finer elements of the surgical scene, such as blood, out of the box. Thus, domain-specific segmentation models have been widely explored for surgical purposes, including instrument segmentation and distinguishing between different types of tissues in the background. Existing methods include U-Net-based models \cite{ronneberger2015u, li2018h, guan2019fully, weng2019unet} which have shown exemplary performance in a wide variety of tasks \cite{siddique2021u} with little training data. This success has been shared across U-net++-based models \cite{zhou2018unet++, zhou2019unet++, lu2021wbc, tulsani2021automated} and DeepLabV3+-based models \cite{chen2018encoder, roy2019segmentation, azad2020attention}. More recently, attention-based mechanisms and transformers have been applied to segmentation tasks - in particular, we point to MANet-based models \cite{fan2020ma, elkarazle2023improved} and SegFormer-based models \cite{xie2021segformer, huang2021missformer, liu2022isegformer}, respectively, for their focus on multi-scale quality on medical images. However, while such models have found success in microvascular challenges, their performance has yet to be shown on macrovascular, surgical data. In this work, we benchmark the performance of popular network architectures UNet, UNet++, DeepLabV3+, MANet, and Segformer on the proposed dataset. 

\section{Methods}
\subsection{Data Collection}

The surgery conducted is a thyroidectomy on a porcine model, which excises the thyroid gland in the anterior neck, performed by the da Vinci \textregistered{} Surgical System (Intuitive Surgical, Inc., Sunnyvale, CA) \cite{ballantyne2003vinci} as shown in Figure \ref{fig1}.
During the thyroidectomy, blood vessels are identified by the operating surgeon and the surgical assistant.
The blood vessels are ruptured by the surgical assistant to induce hemorrhaging.
The operating surgeon is then tasked to stop the bleeding by closing the vessel with electric cautery.
We ensure that the care and use of the animals for data collection are humane and follow federal regulations under IACUC \#S19130 \cite{lin2022semantic}.



We record stereoscopic data through endoscopic camera, specifically the daVinci Surgical System Camera. The data is recorded to the main computer through Robot Operating System (ROS) \cite{ros} at a 1920 by 1080 pixel resolution, later rectified to 640 by 480, at a rate of 30 frames per second --- additional details are shown in Table \ref{tab:datastat}. Annotations are hand-labeled from one camera every 60 frames with the assistance of a polygonal segmentation tool \cite{wada2018labelme}. Each segmentation has two classes, 'blood' and 'background'.

\begin{figure}[t]
    \includegraphics[width=\linewidth]{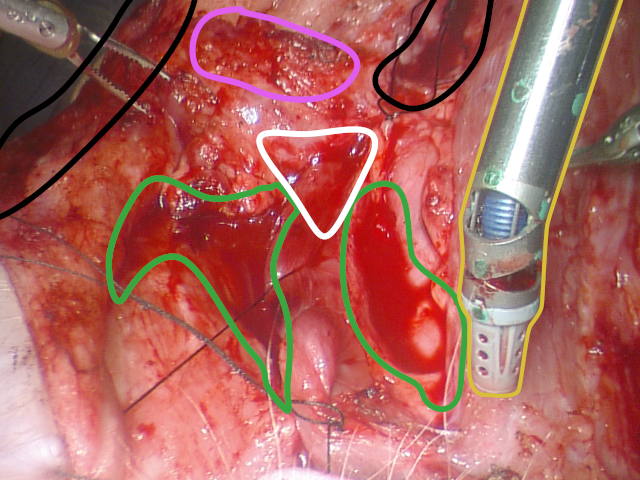}
    \caption{Our image provided to labelers for the annotation comparison. We point out areas to include (green, white) and areas to avoid from blood stains (black, pink) and the suction tool (yellow). With our guidelines, we achieve a STAPLE precision of 0.933 and a specificity of 0.996. }\label{fig:label_guidelines}
\end{figure}

To ensure labeling consistency, we provide our labeling guidelines alongside an annotation comparison. Our annotators were also provided a visual aid, shown in Figure \ref{fig:label_guidelines}.

\begin{enumerate}
    \item Identify and label obvious pools of blood (green), of a depth and area that the suction tool (yellow) can fit into comfortably and perform a majority of the blood suction.
    \item Identify areas of deformation containing lines or groupings of blood that, if slightly deformed, could combine into pools of blood as described in step one. Once identified, determine if the areas in question are close enough to the surgical operation to be labeled (white) or not (black).
    \item Be aware of mislabeling areas that appear to be collections of blood but instead are simply discolorations or stains (pink).
\end{enumerate}

For our comparison, we provide these guidelines to five separate annotators alongside twelve images uniformly selected from our dataset. We then apply the Simultaneous Truth and Performance Level Estimation (STAPLE) algorithm \cite{warfield2004simultaneous}, which computes a probabilistic estimate of the true label and measures the performance level represented by each given label. Through this method, we report an average precision of 0.933 and an average specificity of 0.996 for each annotator as compared to the STAPLE estimate, demonstrating a significant consistency to our labels. 

\begin{figure*}[t!]
    \makebox[1\textwidth]{\includegraphics[width=1\textwidth]{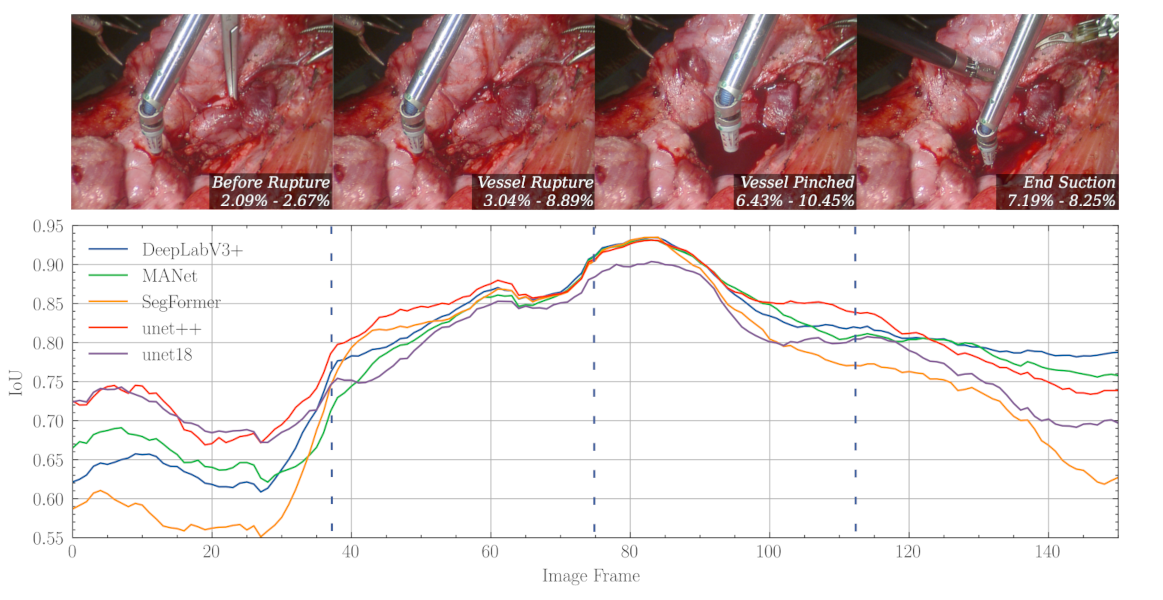}}
    \caption{IoU of each model over trial 9, with a 10-wide running mean window. We also include sample images from each section of the trial, noting the first and third quartiles of the percentages of blood coverage. Although UNet++ performs the best, it is still with difficulty comparable to the other models. We note the performance variance over the dataset, correlated with blood coverage, thus demonstrating the necessity for more nuanced architectures.}\label{fig:frame_IoU}
\end{figure*}

The recorded data present unusual features that distinguish them from previous works in medical image segmentation. Figure \ref{fig:datastat} highlights the blood percentage across the distribution of images. We note a large volume of fluid is distributed on uneven tissue surfaces during operation. The inconsistency of the rupture location and surface causes irregular geometry of blood pools between operations, as opposed to the controlled environment of previous fluid dynamics tasks, making generalization between trials difficult. In addition, the data collection occurs during an active operation, which leads to continuous tool and operator occlusion. The data also feature the blood's wide coverage of the camera frame, with a greatly varying amount of blood per image. This environment best emulates a clinical setting, minimizing the gap between training data and applied space.

\subsection{Benchmark Models}

We benchmark our dataset with five famous segmentation models, which utilize different neural network layers and architecture. We introduce the benchmarking segmentation models and explain the training details.

U-net \cite{ronneberger2015u} is a ubiquitous medical segmentation algorithm, succeeding in numerous endoscopic and extracorporeal tasks. It is a CNN with gradually narrowing layers as an encoder to extract image features followed by a series of upsampling layers as a decoder for precise localization. By including skip connections between these symmetric layers, the network combines semantic features from later in the architecture with the lower-level details of the initial layers.


U-net++ \cite{zhou2018unet++} expands on U-net by redesigning the skip connections between the encoder and decoder to include convolutional layers between them. By upsampling features from deeper levels of the encoder to these convolutional layers, it mimics the effects of the skip connection on a smaller scale. The model also includes deep supervision \cite{lee2015deeply} to enable model pruning.


DeepLabV3+ \cite{chen2018encoder} employs an atrous convolutional layer that feeds into a parallel series of layers of varying kernel sizes, stacking and then compressing to a small feature space. This feature space is then upsampled and convolved with the direct output of the atrous layer before being decoded into the segmentation. This implementation allows for a greater field-of-view during the initial layers, potentially at the cost of low-level localization.


MANet \cite{fan2020ma} is built similarly to U-net, but includes an attention mechanism prior to each upsampling. The first layer to use attention considers spatial dependency across a wide frame, improving semantic inference. Following layers use attention to determine interdependence among features at a high and low level, relying on the upsampled features alongside features from the encoder passed over the network through skip connections.


SegFormer \cite{xie2021segformer} is a recent, popular semantic segmentation algorithm that has a similar encoder-decoder architecture to the previously mentioned models, but is inspired by the recent success of stacked transformer layers as opposed to straightforward convolutional layers. Employing a series of hierarchical transformers to serve as an encoder, then decoding with a lightweight MLP module with skip-connections from the encoder, this design combines many recent successes in image segmentation to achieve new top-of-the-line performance on existing common datasets. We use the PyTorch \cite{paszke2019pytorch} Image Models library implementation for our experiments \cite{rw2019timm}. 


The samples collected and labeled are used to train a variety of state-of-the-art segmentation models \cite{Iakubovskii:2019} on 3 parallel NVIDIA RTX 2080 GPUs. Each model is trained with the rectified images using cross-entropy loss \cite{zhang2018generalized} for 50 epochs, with a batch size of 4. We use the Adam optimizer and an initial learning rate of 0.001, stepping down the learning rate by a factor of 0.1, and prepend each model with an 18-layer residual network (ResNet \cite{he2016deep}) as an encoder pretrained on ImageNet. Finally, we augment our training set through a series of random crops and brightness jitter to encourage generalization, and measure its generalization by ensuring a separation of training and test sets.

\section{Results}

To effectively evaluate the performance of segmentation models, we encompass three different metrics for benchmarking our dataset \cite{taha2015metrics}.

\textbf{Intersection over Union (IoU)} --- IoU is a common metric for evaluating the performance of segmentation, by computing the intersection of the prediction mask and ground-truth mask over the union of the masks.
\begin{equation}
    \textup{IoU}(A, B) = \frac{A \cap B}{A \cup B} 
\end{equation}
where $A$ is the predicted binary mask and $B$ is the ground-truth binary mask.

\textbf{F1 Score (F1)} --- F1 score represents the harmonic mean of the precision and recall:
\begin{equation}
\begin{split}
        \textup{F1}(A, B) & = \frac{2 \cdot \left ( \textup{precision} \cdot \textup{recall} \right )}{\left ( \textup{precision} + \textup{recall} \right )} \\
        \textup{precision} = & \frac{A \cap B}{A}  \qquad  \textup{recall} = \frac{A \cap B}{B}
\end{split}
\end{equation}

\begin{table}[t]
    \centering
    \caption{Each model's average performance and standard deviation across each metric. }
    \label{tab:comparesota}
    \resizebox{\textwidth}{!}{%
        \begin{tabular}{l@{\hskip 0.1cm}|@{\hskip 0.3cm}c@{\hskip 0.3cm}c@{\hskip 0.3cm}c}
            \hline
            Method & IoU & F1 & HD\\
            \hline
            UNet & 0.764(0.072) & 0.864(0.046) & 119.1(52.5)\\
            UNet++ & \textbf{0.793(0.076)} & \textbf{0.883(0.048)} & \textbf{63.3(45.8)}\\
            DeepLabV3+ & 0.780(0.091) & 0.873(0.060) & 79.0(62.4)\\
            MANet & 0.774(0.086) & 0.870(0.056) & 70.1(46.4)\\
            SegFormer & 0.743(0.111) & 0.848(0.075) & 77.3(42.4)\\
            \hline
        \end{tabular}}
\end{table}

\textbf{Hausdorff Distance (HD)} --- The Hausdorff Distance is defined as the greatest distance from a point in one set to the closest point of the other. Hausdorff Distance is an effective metric to evaluate the difference between the shapes of two masks.

\begin{equation}
d_\textup{H}(A, B) = \max \left \{ \sup_{a\epsilon A}d(a, B), \sup_{b\epsilon B}d(A, b) \right \}
\end{equation}

\begin{figure}[t!]
\includegraphics[width=1\linewidth]{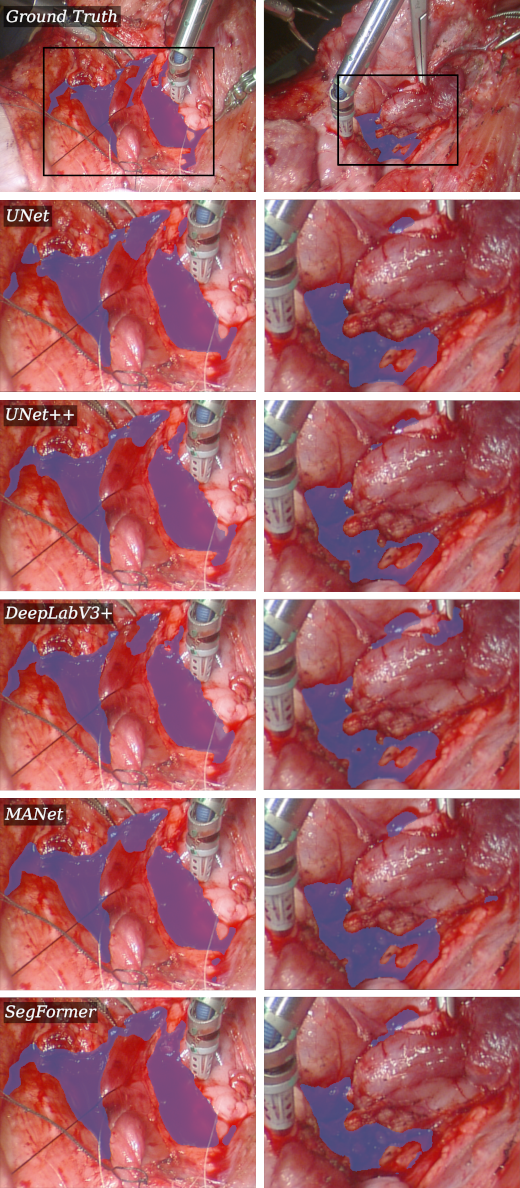}
\caption{From top to bottom: Image w/GT, UNet, UNet++, DeepLabV3+, MANet, Segformer. On images from the 3rd (left) and 9th (right) trials in our dataset.} \label{fig:overlay}
\end{figure}

We report the quantitative results of all three metrics in Table \ref{tab:comparesota}. We also show the qualitative results for selected samples from our dataset in Figure \ref{fig:overlay}.
Due to the fluctuating difficulty of each image in the dataset, the performance of each model varies immensely, reflected in Figure \ref{fig:frame_IoU} and standard deviation in Table \ref{tab:comparesota}. This prevents us from directly comparing between models, so we introduce a series of paired t-tests. Importantly, the performance difference between the best-performing model, UNet++, and the second-best-performing model, DeepLabV3+, is statistically significant with a p-value of 1.03e-5. We likewise perform each of these tests on F1 and Hausdorff Distance, with p-values of 1.84e-7 and 1.29e-2 respectively. In fact, we find that the model performance distributions as ranked by their means are all statistically significant across IoU, with the highest p-value being 2.72e-3.

\begin{figure}[t!]
    \includegraphics[width=\linewidth]{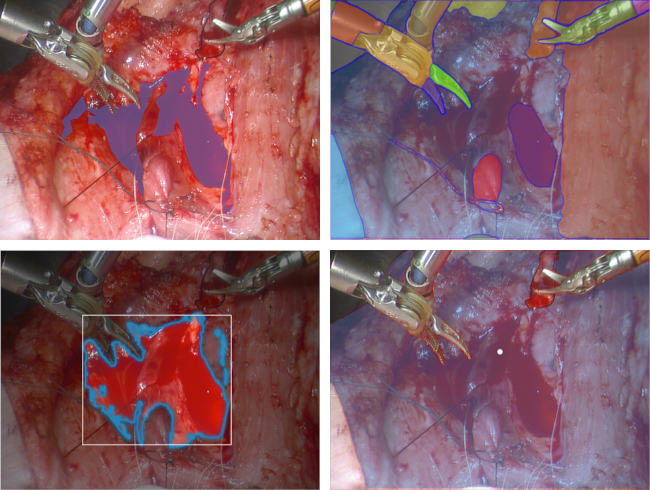}
    \caption{The original image with groundtruth (03-002520, top-left) is segmented with SAM first with no user input (top-right), bounding box (bottom-left) and point (bottom-right). We note that the model is only capable of an approximate outline when the blood takes up a significant proportion of the query space, and even then fails to adequately segment the pool. The Segment Anything model cannot adapt to the surgical scene, demonstrating the necessity for a specialized model.}\label{fig:sam_comparison}
\end{figure}

\section{Discussion}

Given initial comparisons to SAM \cite{kirillov2023segment}, it is clear that SOTA general segmentation algorithms are incapable of capturing pooling geometry to the fidelity necessary for a hemostasis assistant, as shown in Figure \ref{fig:sam_comparison}. The mistaken assignment of the entire surgical scene demonstrates that such networks are unable to adapt to the amorphous and textureless environment. As such, a specialized dataset is necessary to adequately train models for blood segmentation.

HemoSet satisfies this necessity. Blood segmentation is a difficult task to label given the same challenges that we mention for an algorithm --- the glossy and stained surfaces prevent easy annotation across labelers. However, our guidelines have maintained a consistent annotation strategy across our labels, shown by the 0.933 precision metric given by the STAPLE comparison. This consistency establishes the labelers' understanding of the features necessary for blood segmentation that the models do not capture.

Overall, the segmentation masks from our dedicated models capture the general shape of each blood pool, as shown in Figure \ref{fig:overlay}, but suffer from distinct drawbacks that prevent their direct implementation to downstream automation such as an autonomous suction tool or blood loss estimation. For one, each model will often fail to mark thin streams of blood, a clear issue for a segmentation task. In addition, they will mark blood traces or stains rather than pools --- an autonomous suction tool should limit its direct interaction with the tissue under operation as it can slow the suction process, inhibit the primary surgeon, or in rare cases even damage the tissue. The high variation across each model, as depicted in Figure \ref{fig:frame_IoU}, shows the varying difficulties of these challenges in each image.

Our results show UNet++ as the best-performing segmentation algorithm from our experiments. Nevertheless, it still falls short in performance relative to the consistency we measured from our human labelers, and we confidently state that HemoSet presents novel challenges that are not trivially addressed by existing SOTA algorithms. The annotation comparison showed there exist complex features that humans can readily grasp while current segmentation algorithms fail. Our dataset features such challenges prominently to spur the development of improved medical segmentation algorithms.

\section{Conclusion}

With the proposed blood segmentation dataset HemoSet, we provide the first labeled data centering on macrovascular bleeding. The blood pools' erratic geometry and surrounding stained tissue provide a formidable challenge to current segmentation algorithms. By providing a dataset that accurately represents the surgical scene, numerous doors are opened for the community --- estimation of blood loss during surgery, guidance for assistants operating a suction tool, automatic identification of critical bleeding zones, and the development of an autonomous blood suction tool to aid in hemostasis.


In future works, we aim to improve our dataset by expanding its current applications, as well as provide further data with the same qualities we emphasize in mind. Stereoscopic camera data provides the opportunity to explore our dataset with depth imaging, encouraging its use for tracking and reconstruction algorithms. Further iterations may focus on greater variety and volume of blood pools and provide improved comparisons between segmentation models. 




\newpage
\bibliographystyle{IEEEtran}
\bibliography{refs} 
\balance

\end{document}